\documentclass[iop]{emulateapj}
\usepackage{graphicx}
\usepackage{natbib}
\usepackage{amsmath}
\newcommand{\msun}{M$_{\odot}$}

\newcommand{\HI}{H{\sc i}}
\newcommand{\HII}{H{\sc ii}}

\shortauthors{McQuinn et al.}
\shorttitle{Leo~P: How Many Metals can a Very Low-Mass, Isolated Galaxy Retain?}

\begin{document}
\title{Leo~P: How Many Metals can a Very Low-Mass, Isolated Galaxy Retain?\protect\footnotemark[*]}
\footnotetext[*]{Based on observations made with the NASA/ESA Hubble Space Telescope, obtained from the Data Archive at the Space Telescope Science Institute, which is operated by the Association of Universities for Research in Astronomy, Inc., under NASA contract NAS 5-26555.}
\author{Kristen~B.~W. McQuinn\altaffilmark{1,2}, 
Evan D.~Skillman\altaffilmark{1},
Andrew Dolphin\altaffilmark{3},
John M.~Cannon\altaffilmark{4},
John J.~Salzer\altaffilmark{5},
Katherine L.~Rhode\altaffilmark{5},
Elizabeth A.~K.~Adams\altaffilmark{6},
Danielle Berg\altaffilmark{7},
Riccardo Giovanelli\altaffilmark{8},
Martha P.~Haynes\altaffilmark{8}
}

\altaffiltext{1}{Minnesota Institute for Astrophysics, School of Physics and
Astronomy, 116 Church Street, S.E., University of Minnesota,
Minneapolis, MN 55455; \ {\it kmcquinn@astro.umn.edu}}
\altaffiltext{2}{University of Texas at Austin, McDonald Observatory, 2515 Speedway, Stop C1402, Austin, Texas 78712}
\altaffiltext{3}{Raytheon Company, 1151 E. Hermans Road, Tucson, AZ 85756}
\altaffiltext{4}{Department of Physics and Astronomy, 
Macalester College, 1600 Grand Avenue, Saint Paul, MN 55105}
\altaffiltext{5}{Department of Astronomy, Indiana University, 727 East 3rd Street, Bloomington, IN 47405}
\altaffiltext{6}{ASTRON, the Netherlands Institute for Radio Astronomy, Postbus 2, 7990 AA, Dwingeloo, The Netherlands}
\altaffiltext{7}{Center for Gravitation, Cosmology and Astrophysics, Department of Physics, University of Wisconsin Milwaukee, 1900 East Kenwood Boulevard, Milwaukee, WI 53211}
\altaffiltext{8}{Center for Radiophysics and Space Research, Space Sciences Building, Cornell University, Ithaca, NY 14853}

\begin{abstract}
Leo~P is a gas-rich dwarf galaxy with an extremely low gas-phase oxygen abundance (3\% solar). The isolated nature of Leo~P enables a quantitative measurement of metals lost solely due to star formation feedback. We present an inventory of the oxygen atoms in Leo~P based on the gas-phase oxygen abundance measurement, the star formation history, and the chemical enrichment evolution derived from resolved stellar populations. The star formation history also provides the total amount of oxygen produced.  Overall, Leo~P has retained 5\% of its oxygen; 25\% of the retained oxygen is in the stars while 75\% is in the gas phase. This is considerably lower than the $20-25$\% calculated for massive galaxies, supporting the trend for less efficient metal retention for lower mass galaxies. The retention fraction is higher than that calculated for other alpha elements (Mg, Si, Ca) in dSph Milky Way satellites of similar stellar mass and metallicity. Accounting only for the oxygen retained in stars, our results are consistent with those derived for the alpha elements in dSph galaxies. Thus, under the assumption that the dSph galaxies lost the bulk of their gas mass through an environmental process such as tidal stripping, the estimates of retained metal fractions represent underestimates by roughly a factor of four. Because of its isolation, Leo~P provides an important datum for the fraction of metals lost as a function of galaxy mass due to star formation.
\end{abstract} 


\section{Introduction\label{intro}}
The metallicity of a galaxy correlates with stellar mass over a range from 10$^4$ to 10$^{10.5}$ \msun, recognized as the mass-metallicity relation \citep[$M-Z$ relation;][]{Tremonti2004, Berg2012, Kirby2013}. While star formation is responsible for the build-up of metals in a galaxy, measured abundances account for significantly fewer metals in galaxies than expected based on quantified star formation histories (SFHs) and nucleosynthesis yields, particularly at lower masses \citep[e.g.,][and references therein]{Tolstoy2009}. The paucity of metals is generally attributed to three different mechanisms: (i) variations and uncertainties in nucleosynthesis yields, (ii) the dilution of the gas-phase metallicity due to the infall of pristine gas from the outer disk or local extragalactic environment, and/or (iii) the removal of metals through supernova-driven galactic winds. The latter process is the favored explanation of the low-metallicity measurements in dwarf galaxies as the efficiency of the process is predicted to increase with decreasing galaxy mass. 

There have been numerous models quantifying the mass-loss process due to stellar feedback  in low-mass systems \citep[e.g.,][]{Larson1974, Dekel1986, MacLow1999, Ferrara2000}. The prescriptions and details of the models vary, but the premises and overall results are the same: the mechanical luminosity of supernovae, particularly if the supernovae are somewhat spatially and temporally co-located, can evacuate metals from a galaxy while retaining ambient gas. Generally, the metals produced both by the supernovae progenitors and by the supernova process itself are accelerated through the lower-density shocked gas cavity surrounding the supernova. The metals maintain velocities high enough to be pushed out of the host galaxy. In contrast, the ambient gas is denser and cooler, requiring more energy to be accelerated to escape speeds and, thus, remains bound. Two of the seminal works on this process \citep{MacLow1999, Ferrara2000} modeled galaxies with gas masses between $10^6$ \msun $< M_{gas} < 10^9$ \msun\ and found that in this mass range, supernovae were highly efficient in expelling enriched material from a galaxy - either in blow-outs where the bulk of the gas is still bound and may eventually return to the galaxy, or in blow-aways where the bulk of the gas is unbound. These simulations predict that systems massive enough to retain a gaseous disk above a ``threshold'' gas mass of $10^6$ \msun\ will also have a massive enough dark matter halo to retain a minimum amount of metals, which places a predicted lower limit on the oxygen abundance to be $12+\rm{log(O/H)} \sim 7.2$. In the context of chemical evolution, the simulations simplify the analysis as they are able to study a theoretical, isolated, low-mass system without considering environmental effects such as tidal interactions, ram pressure stripping, or gas in-fall. 

Testing the prediction of how many metals may be expelled due to stellar feedback in this low-mass regime has proven difficult. Despite the expectation that low-mass, low-luminosity dwarfs should be the most numerous type of galaxy, there is a bias against detecting such faint systems. Compounding the problem, gas-rich star-forming galaxies are preferentially located outside of group environments, making detection even more challenging at the larger implied distances. Within the Local Group, two star-forming galaxies with a gas mass below the ``threshold'' mass of $10^6$ \msun\ have been cataloged \citep[LGS~3 and Leo~T,][]{McConnachie2012}. However, while spectroscopic measurements of Leo~T's stars have been obtained \citep{Simon2007}, neither LGS~3 nor Leo~T host an \HII\ region. Thus, the gas-phase metallicity in both galaxies is unknown and it is not possible to measure the amount of oxygen retained in the interstellar medium in either system. 

\citet{Kirby2011c} have studied gas-poor Milky Way dwarf spheroidal galaxies (dSphs) in this context. Detailed spectroscopic stellar abundance measurements and star formation history studies of these systems show that the metallicity distribution functions are consistent with the removal of 96\% $->$99\% of the metals. \citet{Kirby2011c} attributed the nearly complete loss of metals to stellar feedback. Yet, given the observational evidence for a morphology-density relation, environmental processes may play a role in the removal of gas and its associated metals.

Through the Arecibo Legacy Fast ALFA survey \citep[ALFALFA,][]{Giovanelli2005, Haynes2011}, an isolated, gas-rich, very low mass galaxy, Leo~P \citep[$M_* = 5.6\times10^5$ \msun;][]{McQuinn2015e}, was discovered \citep{Giovanelli2013} just outside the zero velocity boundary of the Local Group \citep{McQuinn2013}. The galaxy has on-going star formation, an \HII\ region \citep{Rhode2013}, and one of the lowest gas-phase oxygen abundances ever measured \citep[$12+\rm{log(O/H)} = 7.17\pm0.04$;][]{Skillman2013}. While the oxygen abundance is comparable to that of DDO~68 and I~Zw~18, unlike these other galaxies, Leo~P shows no signs of interaction or gas infall \citep{Bernstein-Cooper2014} that could dilute the metallicity. Instead, Leo~P lies on an extension of the $M-Z$ relation to very low luminosities \citep{Berg2012, Skillman2013}. All indications suggest that the evolution of Leo~P is dominated by internal processes, not external events. The serendipitous combination of isolation, relative proximity, gas content, star-forming properties, and the single \HII\ region, makes Leo~P an ideal system in which to test predictions as to whether or not supernovae can expel nearly all the metals from a galaxy in this very low-mass regime. 

We present an oxygen budget for Leo~P, comparing the amount of oxygen produced by star formation over the lifetime of the galaxy with the mass of oxygen currently contained in the gas and stars. We discuss the deficit in oxygen in Leo~P based on this accounting and the implications for metal-loss in low-mass galaxies due to star formation feedback versus other mechanisms. 

%tab:properties
%McQuinn et al. 2015 Leo P: Metals Retained Table 1
\begin{table}
\begin{center}
\caption{Properties of Leo~P}
\label{tab:properties}
\end{center}
\begin{center}
\vspace{-5pt}
\begin{tabular}{lr}
\hline 
\hline 
R.A. (J2000)            & 10:21:45.1                    \\
Decl. (J2000)           & $+$18:05:17   \\
Distance (Mpc)          & 1.62$\pm0.15$ \\
$M_V$ (mag)             & $-9.27\pm0.20$        \\
$M_*$ (\msun)           & $5.6\times10^5$    \\
$M_{HI}$ (\msun)        & $8.1\times10^5$    \\
$12+$log(O/H)           & 7.17$\pm0.04$         \\
\hline
\end{tabular}
\end{center}
\textbf{References:} \citet{McQuinn2015e, Giovanelli2013, Skillman2013}.
\\
\end{table}

\section{The Observations and Star Formation History of Leo~P\label{data}}
Hubble Space Telescope ($HST$) optical imaging of Leo~P was obtained over 17 orbits (GO 13376; PI McQuinn). A full description of the observations, data reduction, and methodology for reconstructing the SFH is described in \citet{McQuinn2015e}. The resulting color-magnitude diagram (CMD) reaches $\sim2$ mag below the red clump with a clearly defined main sequence, red giant branch, red clump, horizontal branch, and sparsely populated helium burning sequences. 

\begin{figure*}[ht]
\includegraphics[width=\textwidth]{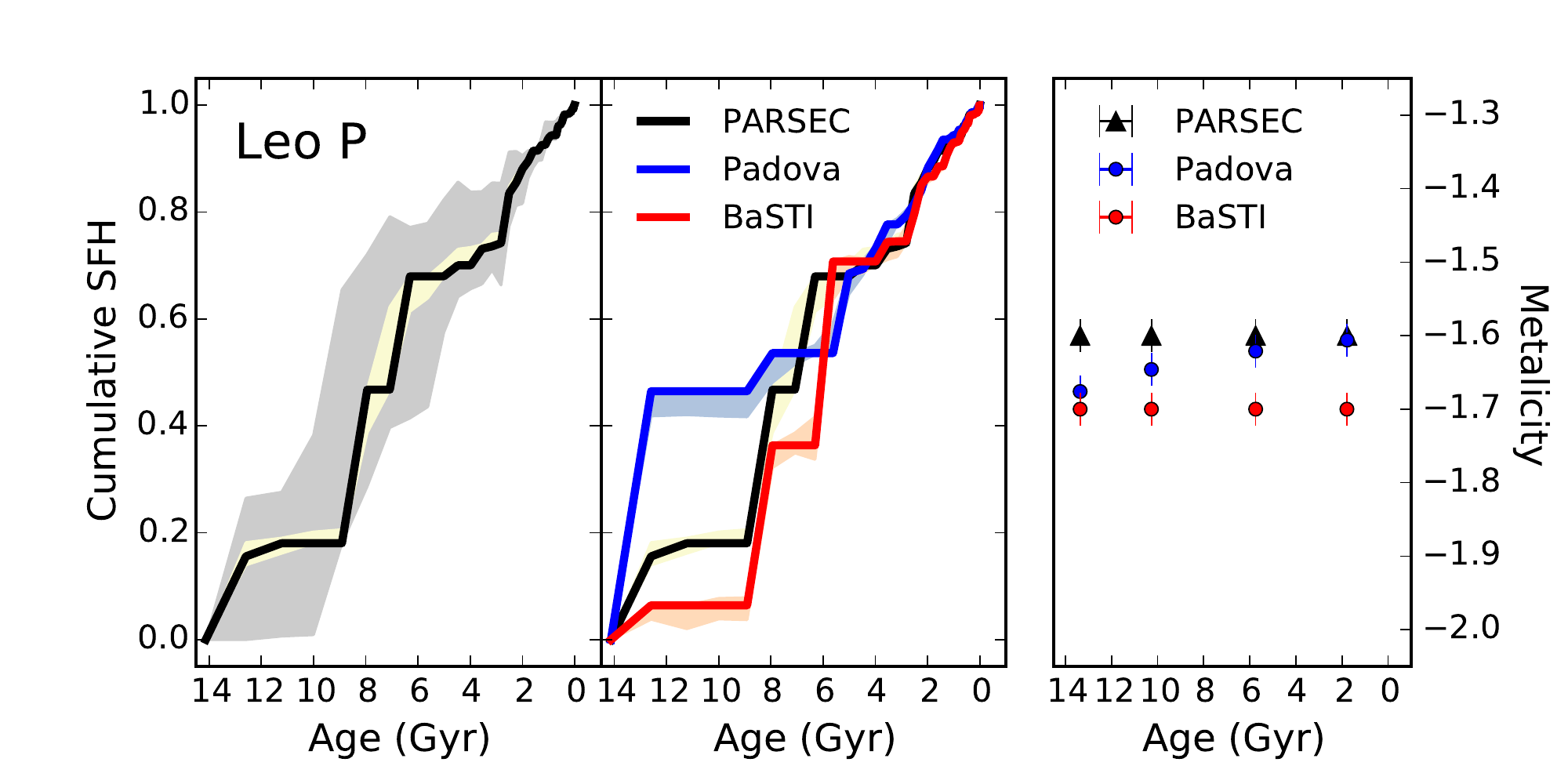}
\caption{\textit{Left panel:} Best-fitting cumulative SFH for Leo~P. Random uncertainties due to the finite numbers of stars in the CMD are plotted in yellow, combined systematic and random uncertainties are plotted in grey.  \textit{Middle panel:} The best-fitting cumulative SFH for Leo~P using three different stellar libraries; random uncertainties are plotted as shaded regions. \textit{Right panel:} Chemical evolution [M/H] solution for Leo~P derived using three different stellar libraries including random uncertainties. For clarity, we plot limited modeled values of stellar metallicity during periods of star formation. The full chemical enrichment solution is consistent with the points shown, with little increase in metallicity across all ages. Figure reproduced from \citet{McQuinn2015e}.}
\label{fig:sfh_metals}
\end{figure*}

The SFH was measured using the numerical CMD fitting program {\tt MATCH} \citep{Dolphin2002a}, as described in \citet{McQuinn2015e}. The SFH solution is based on a Kroupa IMF \citep{Kroupa2001}, and an assumed binary fraction of 35\% with a flat secondary distribution. We used three different stellar evolution models including the newly released Padova-Trieste models \citep[PARSEC;][]{Bressan2012, Chen2014, Tang2014}, the Padua models \citep{Marigo2008} with updated AGB tracks from \citet{Girardi2010}, and the BaSTI models \citep{Pietrinferni2004}. The three models allow us to probe how dependent the best-fitting SFH solution is on the stellar model used. The chemical enrichment history, $Z(t)$, was assumed to be a continuous, non-decreasing function. Uncertainties on the SFHs take into account both systematic uncertainties from the stellar evolution models \citep{Dolphin2012} and random uncertainties due to the finite number of stars in a CMD \citep{Dolphin2013}. 

Figure~\ref{fig:sfh_metals} plots the cumulative SFHs derived using three stellar evolution libraries and the chemical evolution of Leo~P from \citet{McQuinn2015e} based on the best-fitting modeled CMD from {\tt MATCH}. Most relevant for this work, the right panel in Figure~\ref{fig:sfh_metals} shows the chemical evolution history from all three libraries. There is only a modest difference in stellar metallicity from [M/H] $=-1.7$ to $-1.6$ between the libraries, suggesting that our subsequent analysis using the chemical evolution history to trace stellar metallicity is independent of the choice of model. 

\section{The Oxygen Budget in Leo~P\label{metals}}
We inventory the oxygen atoms in Leo~P by calculating the mass of oxygen produced by stars on one side of the ledger and the oxygen atoms retained in stars and gas on the other side of the ledger. The calculations are based on a number of assumptions including (i) oxygen that is recycled into the ISM is uniformly mixed (i.e., the gas-phase oxygen abundance measurement from the single \HII\ region is representative of the ISM for the galaxy), (ii) the IMF is constant, and (iii) 30\% of the total stellar mass formed is recycled to the ISM \citep[the oxygen recycling fraction, ``R'', see][]{Kennicutt1994}. 

\subsection{Total Oxygen Produced by Star Formation}
The total amount of oxygen produced by star formation can be estimated by calculating the total mass of stars formed over the lifetime of the galaxy from the SFR(t) in Figure~\ref{fig:sfh_metals} and assuming a constant mass of newly synthesized oxygen per solar mass of gas converted to stars (P).

\begin{equation}
M^O_{total} = P * \int_{t_i}^{t_f} SFR(t)~dt. \label{eq:O_total}
\end{equation}

Integrating the SFR(t) over time yields a total stellar mass $= 8.6\pm0.2\times10^5$ \msun; uncertainties are based on the finite number of stars in the CMD. We include systematic uncertainties in our final numbers below. We use a constant value for P of 0.00658 from \citet{Nomoto2006} to be consistent with the similar calculations for dSphs from \citet{Kirby2011c}. Note that the value of P depends on models of nucleosynthesis yields and the IMF. Our choice of P falls in the middle of the range from $0.0054 - 0.0086$, which brackets yields calculated based on different models of nucleosynthesis yields and IMFs used by \citet{Zahid2012} in a statistical analysis of the O budget in more massive galaxies from SDSS data. From the above integration, the total mass of oxygen produced by star formation in Leo~P is $5.7\pm0.1\times10^3$ \msun. 

\subsection{Oxygen in the Gas}
Here we consider only the oxygen measured in the gas-phase in Leo~P. As listed in Table~\ref{tab:properties}, $M_{HI} = 8.1\times10^5$ \msun, while the $M_{H_2}$ is still uncertain. High-sensitivity observations with CARMA did not detect CO in Leo~P, instead placing upper limits on the CO luminosity \citep{Warren2015}. The non-detection coupled with the uncertain ratio of CO to H$_2$ at low metallicities makes it difficult to quantify the amount of hydrogen in molecular form in the galaxy \citep[e.g.,][and references therein]{Bolatto2013}. Observational studies of dwarf galaxies suggest that the molecular hydrogen content is a small percentage of the total gas mass \citep{Leroy2008, McQuinn2012b}. Thus, we explicitly assume the H content in molecular form is negligible. Additionally, because dust content scales with metallicity and is known to be negligible at the metallicity of Leo~P \citep[e.g.,][]{Cannon2002}, we assume the O content in the dust is also negligible.

The mass of oxygen in the gas-phase of Leo~P can be estimated based on the \HII\ region abundance measurement:

\begin{equation}
M^O_{gas} = 10^{\rm{log(O/H)}} \times ({m_O/m_p}) \times M_{HI}. \label{eq:O_gas}\\
\end{equation}

\noindent Using the measured gas-phase oxygen abundance in Leo~P, this becomes:
\begin{equation}
M^O_{gas} = 10^{-4.83} \times 16 \times 8.1 \times 10^5. \label{eq:O_gas_numbers}\\
\end{equation}

Therefore, the total mass of oxygen in the gas-phase in Leo~P is $M^O_{gas} = 192\pm18$ \msun. 
Observations of \HII\ region abundances have shown that metallicity gradients and variations are small to non-existent in low-mass galaxies \citep{Skillman1989, Kobulnicky1996, Lee2006, Croxall2009, Berg2012}. Although one might expect significant spatial variations in ISM abundances due to local pollution, these are not observed, and the current understanding is that newly formed elements are released into the hot phase of the ISM and mix back into the cooler phases of the ISM in a uniform manner over longer timescales \citep{Tenorio-Tagle1996, Kobulnicky1997, Legrand2000}. Thus, the assumption that the gas-phase O abundance value from a single \HII\ region is representative of the average in the ISM of Leo P is well supported. This gas-phase abundance is slightly higher than the stellar metallicity constraints from the CMD-fitting technique, providing further support that this assumption is reasonable.

%tab:o_balance
%McQuinn et al. 2015 Leo P: Metals Retained Table 2
\begin{table}
\begin{center}
\caption{Oxygen Budget in Leo~P}
\label{tab:o_balance}
\end{center}
\begin{center}
\vspace{-15pt}
\begin{tabular}{lr}
\hline 
\hline 
Total O Produced                   	& $5.7\pm0.1\times10^3$ \msun\\
O locked in stars (O$_{*}$ )    	& $68\pm16$ \msun           	\\
O in the ISM (O$_{gas}$)         	& 192$\pm18$ \msun             			\\
``Missing'' O ($\Delta$O)         	& $5.4\pm0.1\times10^3$ \msun\\
Percent of O Lost                      	& 95$\pm2$\%		                  \\
\hline
\end{tabular}
\end{center}
\tablecomments{Inventory of the oxygen atoms in Leo~P based on the star formation history, chemical evolution history, and the gas-phase oxygen abundance measurement. Uncertainties on the individual O amounts are based on the finite number of stars in the CMD; final uncertainties on the percent of O lost include both random and systematic uncertainties. Despite its extremely low-mass, Leo~P has retained 5\% of its oxygen.}
\end{table}

\subsection{Oxygen Locked in Stars}
The oxygen locked in the stars can be estimated using the SFR(t) and metallicity evolution derived from the CMD fitting results (see Figure~\ref{fig:sfh_metals}). In the absence of stellar spectroscopic measurements, this technique provides an alternative independent approach to estimating the oxygen content in the stars. Assuming an solar oxygen abundance of $12+$log(O/H) $= 8.69$ \citep{Asplund2009}, we multiply the oxygen enrichment as a function of time by the mass of stars formed over each time period, and assume an oxygen recycling fraction (R) of 30\%:

\begin{equation}
M^O_{*} = (1 - R) * \sum (SFR(t) ~ \delta t ~ Z^O_{stars}), \label{eq:O_stars}
\end{equation}

\noindent where:
\begin{equation}
\rm{log (Z^O_{stars})} = \rm{log(O/H)} + \rm{log}\frac{m_O/m_p}{X * m_H + Y * m_{He}}. \end{equation}

\noindent Assuming X and Y to be 75\% and 25\% respectively and the chemical evolution history from Figure~\ref{fig:sfh_metals}, this yields an estimate of the total mass of oxygen locked in stars to be $M^O_* = 68\pm16$ \msun. Accounting for the distribution of mass between the gas and the stars, the ratio of $M^O_{gas}/M_{gas}$ to $M^O_*/M_*$ is $2:1$. In other words, there are twice as many O atoms in the gas per unit mass than in the stars.

\subsection{The Deficit of Oxygen in Leo~P}
The total oxygen produced by star formation is significantly greater than the combined mass of oxygen in the gas and stars. The deficit in oxygen can be found by the simple calculation:

\begin{equation}
\Delta M^O = M^O_{total} - M^O_{gas} - M^O_*, \label{eq:O_deficit}
\end{equation}

\noindent which is $5.4\pm0.1\times10^3$ \msun. Systematic uncertainties were measured via Monte Carlo simulations by varying the observed CMD in both luminosity and temperature, and re-solving for the amount of O lost. Including these uncertainties, we find $95\pm2$\% of the oxygen has been expelled from the galaxy.  The final numbers are listed in Table~\ref{tab:o_balance}. The same calculations performed using the Padova and BaSTI models both yield values of 95$\pm2$\% for the fractional loss of O, consistent with our adopted value from the PARSEC models.

\section{Discussion\label{conclusions}}
Based on the SFH, chemical evolution history, and gas-phase oxygen abundance measurements, we have estimated that Leo~P has lost 95\% of its oxygen. The retention of 5\% is still consistent with the almost total lack of chemical enrichment implied from the three SFH solutions. Comparing the relative amounts of oxygen in the stellar and gas phases to that produced by star formation implies that one quarter of the oxygen retained is locked in the stars, while three quarters is retained in the gas. This demonstrates the importance of performing an inventory of oxygen in this low-mass regime in a galaxy that has not had its gas content removed.

\begin{figure}[ht]
\includegraphics[width=0.5\textwidth]{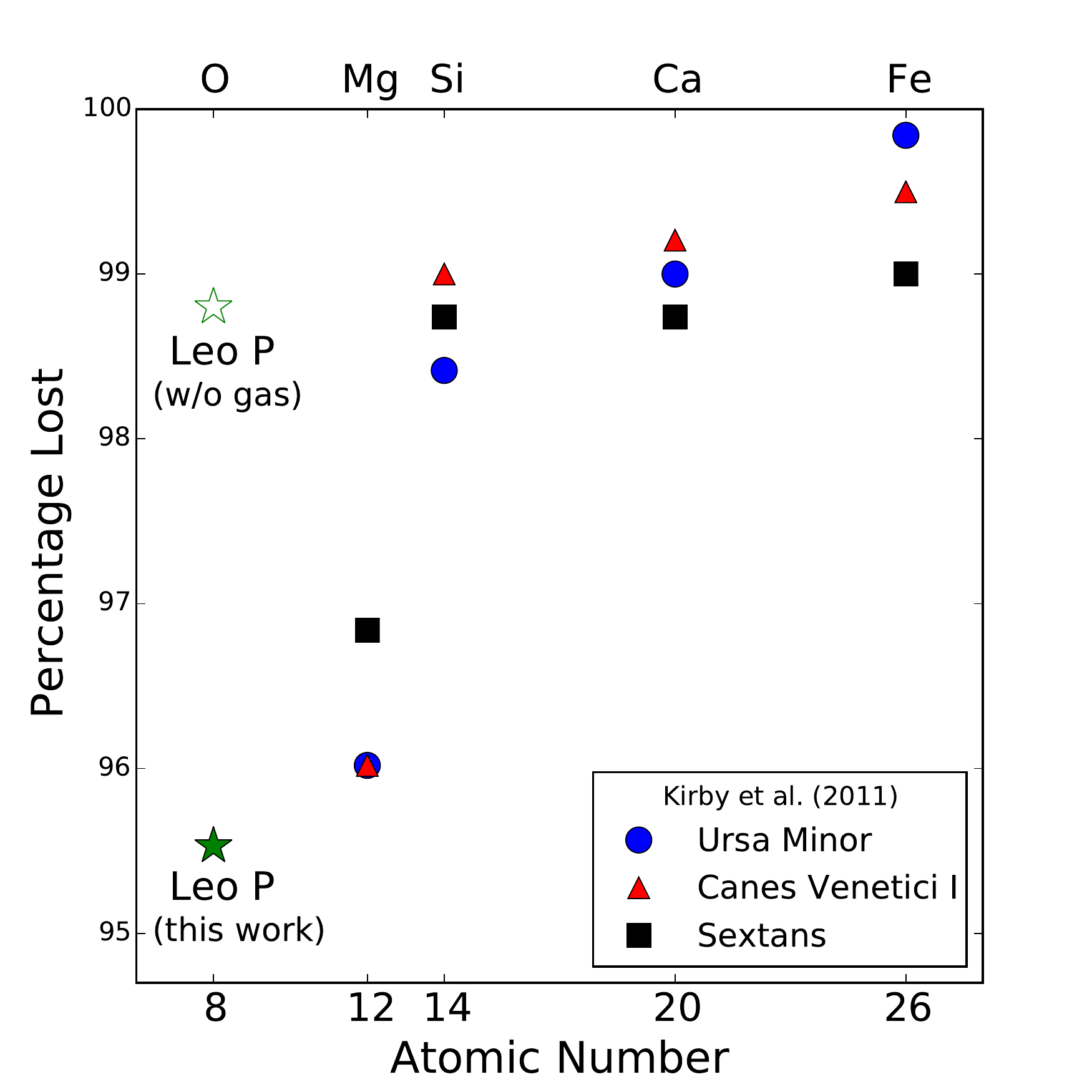}
\caption{The percentage of  O lost in Leo~P compared to the percentage of metals lost in 3 Milky Way dSphs of comparable stellar mass from \citet{Kirby2011c}. Leo~P has retained its gas content and thus a larger fraction of O. If the O atoms in the gas phase are ignored, the percentage of O lost is comparable to gas-poor dSphs (unfilled star point). Also note the discrepant low value for Mg loss in the dSphs.}
\label{fig:percent_lost}
\end{figure}

Figure~\ref{fig:percent_lost} presents a comparison of the percentage of O lost in Leo~P compared with the percentages of four other metal species lost from  similarly low mass dSphs \citep[Ursa Minor, Canes Venetici~I, Sextans;][]{Kirby2011c}. As seen in the figure, Leo~P has retained a larger fraction of metals than these dSphs. 
Presumably due to its isolation, Leo~P has retained a significant amount of gas and, thus, retained a larger fraction of metals. Indeed, three quarters of the O content in Leo~P is in the gas.  Note that considering only the oxygen locked up in stars in Leo~P yields a loss fraction of $\sim99$\%, consistent with the results for the dSphs.  If the dSphs lost a significant amount of their gas due to tidal and/or ram pressure stripping, including the metals present in the gas at the time of removal, then the estimates from \citet{Kirby2011c} represent upper limits to the metals lost due to stellar feedback. 
That is, calculations which do not take into consideration the retention of metals in the gas phase will underestimate the ability of dwarf galaxies to retain metals. 

Also note in Figure~\ref{fig:percent_lost} that the percentage of Mg lost is different from the other alpha elements, Si and Ca; for a constant IMF, these elements are produced at a constant ratio, and presumably would be evacuated from the galaxies by the same processes (i.e., galactic outflows and stripping mechanisms). While we highlight only three dwarf satellites that are of similar stellar mass to Leo~P, we note that this trend is seen in the larger sample of galaxies in \citet{Kirby2011c}. It is not clear why Mg has a different retention fraction.

The models of \citet{MacLow1999} and \citet{Ferrara2000} predict that below $M_{gas} = 10^7$ \msun, the metal ejection efficiency reaches 100\% across the entire range probed of mechanical luminosities due to stellar feedback. However, as noted in these studies, the simulations were designed with extreme conditions. The energy from supernovae is assumed to be injected {\it instantaneously} into the interstellar medium in the central 100 pc of a galaxy. Stellar feedback occurring on such  a short timescale and confined to such a small spatial region would be more effective at driving outflows in a low-mass galaxy than in more realistic conditions \citep[e.g.,][]{Strickland1999}. 

In Leo~P, the star formation occurs over extended periods of time and is distributed across the galaxy. The  oxygen retained in Leo~P is what may be expected without the a priori temporal and spatial star formation assumptions.  Leo~P  allows a direct measure of the role supernovae play in determining the metallicity of a galaxy below the gas mass threshold of $10^6$ \msun. The low oxygen abundance in the present day ISM of Leo P can be explained by a range of different chemical evolution histories, including ones that include dilution by gas accretion at early times. However, because we account for the total mass of oxygen produced versus the amount in the gas and stars, the missing 95\% of oxygen atoms cannot be explained by dilution. This supports the results from \citet{Bernstein-Cooper2014} who concluded that the low metallicity was likely not due to the infall of pristine gas based on the relatively ordered \HI\ kinematics. Note also that \citet{Bernstein-Cooper2014} estimate that the ratio of total mass to baryonic mass in Leo~P is $\sim$ 15:1. Since this is only twice the universal dark matter to baryonic matter ratio, this means that Leo~P has lost, at most, only half of its baryons to processes like supernova feedback. 

The 5\% of oxygen retained in Leo~P is considerably lower than estimates of $20-25$\% of metals retained in more massive galaxies \citep[$10^9$ \msun $< M_* < 10^{11.5}$ \msun;][]{Peeples2014}, with the expected result that the retention fraction of metals decreases with decreasing mass for lower mass galaxies. 

While stellar feedback is likely responsible for the paucity of metals in Leo~P, it is still speculative whether the missing oxygen atoms have escaped the potential well of the galaxy completely or if they reside in a halo of hot coronal gas. As more measurements become available, we will be able to better quantify the fraction of metals and gas lost due to star formation as a function of galaxy mass.

{\bf Acknowledgments}
This work was supported by NASA through grant GO-13376 from the Space Telescope Institute, which is operated by Aura, Inc., under NASA contract NAS5-26555. J.~M.~C. is supported by NSF grant AST$-$1211683. The authors would like to thank NED for including Leo~P in the database and M. Peeples for insightful discussions. The authors would also like to thank the anonymous referee for helpful comments.

{\it Facilities:} \facility{Hubble Space Telescope}

\end{document}